\documentclass[pra,twocolumn,showpacs]{revtex4}
\usepackage{graphicx} 
\usepackage{amsmath} 
\usepackage{amsfonts} 
\usepackage{mathrsfs} 
\newcommand{\Hb}{{\hat{\mathcal H}}_{\beta}}

\newcommand{\ep}{\epsilon} 
\newcommand{\ps}{\ps}

\newcommand{\Ih}{{\hat I}}

\newcommand{\Up}{{\hat{\mathcal U}}}

\newcommand{\Np}{{\hat{\mathcal N}}}

\newcommand{\Ub}{\Up_{\beta}}

\begin{document} 
\title{Experimental verification of a one-parameter scaling law  
for the quantum and ``classical'' resonances  
of the atom-optics kicked rotor} 
\author{Sandro Wimberger,$^{1}$ Mark Sadgrove,$^2$  
Scott Parkins,$^2$ Rainer Leonhardt$^2$} 
\affiliation{ 
$^1$Dipartimento di Fisica E. Fermi, Universit\`{a} di Pisa, 
Largo Pontecorvo 3, 56127 Pisa, Italy \\ 
$^2$Department of Physics,  
University of Auckland, Private Bag 92019, Auckland, New Zealand} 
 
\begin{abstract} 
We present experimental measurements of the mean energy in the 
vicinity of the first and second 
quantum resonances of the atom optics kicked rotor
 for a number of different  
experimental parameters. Our data is rescaled and compared 
with the one parameter $\epsilon$--classical scaling function developed 
to describe the quantum resonance peaks. 
Additionally, experimental data is presented for the  ``classical'' 
resonance which occurs in the limit as the kicking period goes to  
zero. This resonance is found to be analogous to the quantum resonances, 
and a similar one-parameter classical scaling function is derived, and
found to match our experimental results. The  
width of the quantum and classical resonance peaks is compared, 
and their Sub-Fourier nature examined. 
\end{abstract} 
 
\maketitle 
\date{today} 
 
\section{Introduction} 
\label{intro} 
 
The heart of experimentally testing and controlling classical and quantum 
systems often lies in the introduction of an external periodic driving 
force \cite{LL92,Bayfield,Dem}.  
The driving probes system specific properties, the knowledge of which  
allows one, in turn, to 
understand and to optimally control the system at hand. 
In particular, driven systems often exhibit resonance like behavior if the  
external driving frequency matches the natural frequency of the unperturbed 
system.  
 
Typical nonlinear classical systems are resonant  
for only a finite interaction time since the 
driving itself forces the system to gain energy and hence drift  
out of resonance. 
Only if the natural frequencies are independent of the energy as for the 
linear (harmonic) oscillator, can the system absorb energy on resonance 
indefinitely. 
In the quantum world, the situation may be different by virtue of the  
unperturbed system possibly having a 
discrete energy spectrum. If this spectrum 
shows an appropriate scaling in the excitation quantum number, resonant 
motion can persist forever.  
 
A simple example of such a system is provided 
by the free rotor, whose energy spectrum scales quadratically in the 
excitation quantum number (due to periodic boundary conditions for the motion 
on the circle). Kicking the rotor periodically in time with a frequency  
commensurable with the energy difference of two neighboring levels leads to 
perfectly resonant driving. These so called quantum resonances of the 
well-studied kicked rotor (KR) \cite{Casati} have been known theoretically 
for some time \cite{Izr}, but the first traces of this example of 
frequency-matched driving have only recently come to light in experiments 
with cold atoms \cite{QRexp,QRexpnoise}. 
Such experiments \cite{QRexpnoise} and theoretical studies \cite{sandro,Pisa}
have also shown the surprisingly 
robust nature of these resonances in the presence of noise and perturbations.
 
Experimentally, the quantum resonances of the KR are hard to detect 
for two principle reasons. Firstly, only a relatively small proportion of
atoms are kicked resonantly for the following reason:
ideally, the atomic motion is along a line, which 
introduces an additional parameter, namely the non-integer part of the 
atomic momentum, i.e. the atom's quasi-momentum. Treating the atoms 
independently, their motion can be mapped onto the circle owing to the spatial 
periodicity of the standing wave, which makes the quasi-momentum a constant of 
the motion. However, only some values of 
quasi-momentum allow resonant driving to occur \cite{Izr}. All
other values induce a dephasing in the evolution which 
hinders the resonant kicking of the atoms
(see Section \ref{theory} for  
details). Secondly, if an atom is kicked resonantly it 
moves extremely quickly; 
in fact its energy grows quadratically in time (so-called ballistic  
propagation). 
These fast atoms quickly escape any fixed experimental detection window 
after a sufficiently large number of kicks \cite{QRexp,QRexpnoise}.  
 
In this paper, we report experimental data which shows the behavior of  
a typical experimental ensemble of cold atoms under resonant driving.  
Our main observable is the mean energy of the atomic ensemble  measured 
after a fixed number of kicks and scanned over the resonant kicking  
frequency or period. We verify a recently derived 
single-parameter scaling law  
of the resonant peak seen when scanning the energy vs. the period  
\cite{WGF,WGF2,sandro}. The scaling 
law allows us, for the first time, to clearly resolve the resonance peak 
structure because it reduces the dynamics to a {\em stationary} and 
experimentally robust signature of the quantum resonant motion. 

\begin{figure*} 
\centering 
\includegraphics[height=7cm]{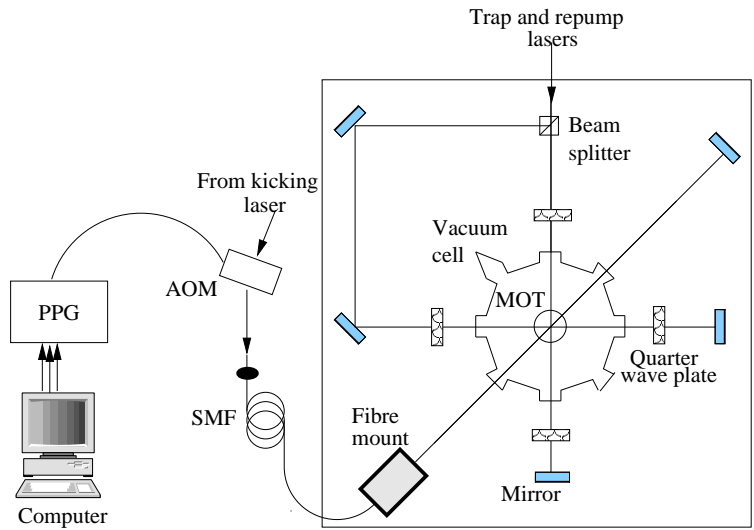} 
\caption{\label{fig:lab} Schematic diagram of our experimental set up. A  
standard six beam magneto optical trap (MOT) of about $10^5$ Cs atoms 
is formed inside a vacuum cell at the intersection 
of 3 retroreflected ``trapping'' beams (vertical beams and (anti) Helmholtz  
coils are not shown).
A standing wave is formed across the cloud of 
atoms by retroreflecting light from a ``kicking laser'', which is  
transported to the MOT by means of a single mode fibre (SMF). This light 
is pulsed on and off by an acousto--optic modulator (AOM)  
which is gated by a programmable pulse generator (PPG). The PPG's pulse 
train is uploaded from a computer, which also controls the timing  
of the experiment (e.g. when the trapping AOM and anti--Helmholtz coils 
are turned on and off).  
} 
\end{figure*} 
 
After a short review of our experimental setup in Section \ref{setup} and the
theoretical treatment of the atom-optics kicked 
rotor close to quantum resonance in Section \ref{theory}, we present 
experimental data for the mean energies around the first  
two fundamental quantum resonances of the kicked atom. From this data, we 
extract the afore mentioned scaling law in Section \ref{sthree}. The effect 
of the quasi-momentum (as a typical quantum variable) on the motion 
disappears in the classical limit of the kicked rotor, when the kicking period 
approaches zero \cite{Izr,Fish}. In the latter case, the rotor is constantly 
driven, and a ballistic motion occurs for {\em all} members of the atomic 
ensemble \cite{we}. Both phenomena, the quantum and the ``classical'' 
(for vanishing kicking period) resonance are related to one another by  
a purely classical theory developed previously for the quantum resonance 
peaks \cite{WGF,WGF2,sandro}. 
 
In Section \ref{sfour} we focus on the first direct comparison of the behavior 
of the ensemble averaged energies in the case of the ``classical'' and the 
quantum resonance. In particular, the Sub-Fourier scaling of the resonance 
peaks in the mean energy as a function of the kick number is discussed. 
The latter makes both types of resonances studied here a potential source 
of high-precision measurements of system specific parameters.

 
\section{Experimental setup}
\label{setup}

Our experimental system is a realization of the paradigmatic kicked rotor 
(KR) model \cite{GSZ,Moore},  
whose relevance lies in the fact that it shows the basic features 
of a complex dynamical system, and it may be used to locally (in energy) 
approximate much more complicated systems, such as microwave-driven Rydberg 
atoms \cite{IEEE}, or an ion in a particle accelerator \cite{Chirikov,LL92}. 
 
Our experiments utilise a cloud of about $10^5$ cold Caesium atoms, provided  
by a standard six beam magneto--optical trap (MOT) \cite{Monroe1990}. 
The typical momentum spread of the atomic sample lies between 
4 and 8 two--photon recoils. The shape of the initial momentum distribution
is well approximated by a Gaussian with standard deviation $\sigma_p 
\simeq (4-8)\times 2\hbar k_L$,
centered at zero momentum \cite{Raizen}, 
although significant non--Gaussian tails
can exist \cite{we}. 
The width is measured in units of
two--photon recoils, corresponding to the wavelength of the kicking laser
$\lambda_L = 2\pi/k_L$. The fractional parts in these units
of the initial momenta,
i.e. the quasi-momentum discussed below, are practically uniformly
distributed in the fundamental Brillouin zone defined by the
periodic kick potential \cite{WGF}.


As shown in Fig. \ref{fig:lab}, the atoms interact  
with a pulsed, far-detuned optical standing wave which is created by  
retroreflecting the light from a 150mW (slave) diode laser which is 
injection locked to a lower power (master) diode laser at a wavelength 
of $\lambda_L=852 \; \rm nm$. 
Power fluctuations were minimal during the experiments  
performed here ($\sim 1\%$) although larger drifts occurred over the course 
of many experimental runs. 
Accurate pulse timing is achieved using a custom built programmable 
pulse generator (PPG) to gate an acousto--optic modulator. The PPG 
is programmed by a computer running the  $\rm RTLinux^{\tiny \texttrademark}$  
operating system kernel \cite{RTLinuxFAQ} which controls the timing of the 
experimental sequence (aside from the pulse train itself).  
Experimentally, we approximate $\delta$-kicks by pulses of width $\tau_p$  
which are approximately rectangular in shape. The lowest value of $\tau_p$ 
used in our experiments was $240 \; \rm ns$ and the highest was 
$480 \; \rm ns$. For the experiments reported here, the effect of the finite
width of the kicking pulses \cite{Raizen,Fishman} turns out 
to be negligible, since fairly small numbers of kicks (less than 
$20$) and low kicking strengths are used. In the case
where the $\tau \rightarrow 0$ limit is being investigated experimentally, 
the $\delta$--kick assumption is clearly not valid \cite{we, Sadgrove2005}. 
This restricts us to a minimum period $\tau = 330 \; \rm ns$, for
$\tau_p = 240 \; \rm ns$, in our study of the ``classical'' resonance
peaks.

In a typical experimental run, the cooled atoms were released from the  
MOT and subjected to up to 16 standing wave pulses, then allowed to 
expand for an additional free drift time in order to resolve 
the atomic momenta. After this expansion time, the trapping beam is  
switched on and the atoms are frozen in space by optical mollases. 
A CCD image of the resulting fluorescence is recorded and used to 
infer the momentum distribution of the atoms using standard time  
of flight techniques \cite{QRexp}. The mean energy of the  
atomic ensemble may then be inferred by calculating the 
second moment of the experimental momentum distribution. 

Kicking laser powers of up to $30\; \rm mW$ were employed, and detunings from
the  $6S_{1/2}(F=4) \rightarrow 6P_{3/2}(F'=5)$ transition of Cesium
of $500 \;\rm MHz$ and 
$1\;\rm GHz$ were used for the classical and quantum resonance
scans respectively. These parameters produced spontaneous emission rates of
$< 0.5\%$ per kick for the quantum resonance scans, which was low enough to ensure 
that the structure of the peaks was not 
affected for the low kick numbers used here.

\section{$\epsilon$-classical dynamics near the
fundamental quantum resonances}
\label{theory} 

We now consider the theoretical treatment of the atom-optics kicked rotor
near quantum resonance. 
The Hamiltonian that generates the time evolution of the atomic wave function 
is (in dimensionless form) \cite{GSZ,QRexp} 
\begin{equation} 
H(t') =\frac{p^2}{2} + k\cos(z)\sum_{t=0}^{N} 
 \delta (t'-t\tau)\;, 
\label{eq:ham} 
\end{equation} 
where $p$ is the atomic momentum in units of $2\hbar k_L$  
(i.e. of two-photon recoils), $z$ is the atomic position in units of  
$2k_L$, $t'$ is time and $t$ is an integer which counts the kicks. 
In our units, the kicking period $\tau$ may also be viewed as a scaled Planck constant  
as defined by the equation $\tau=8E_R T/\hbar$, 
where $E_R=\hbar^2 k_L^2/2M$ is the recoil energy (associated with 
the energy change of a Caesium atom of mass $M$
after emission of a photon of wavelength  
$\lambda_L = 2\pi/k_L = 852\;\rm$nm). The dimensionless parameter $k$  
is the kicking strength of the system and is proportional to the kicking laser
intensity.  

\begin{figure*} 
\centering 
\includegraphics[width=10cm]{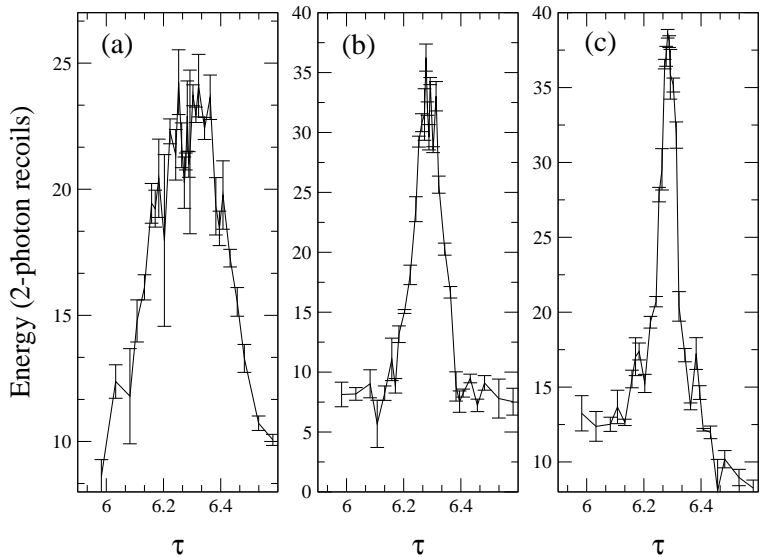} 
\caption{\label{fig:qr2pi} 
Experimentally measured mean energies around  
the first quantum resonance at $\tau=2\pi$ after (a) 5, (b) 10 and (c) 15 
kicks. Error bars show an average over three independent experiments. 
The kicking strength and  
initial momentum standard deviation were measured to be 
$k=4.1 \pm 0.6$ and $\sigma_p=5.9\pm0.2$ respectively. Note that the  
estimated errors in these parameters do not take into account systematic  
drifts which take place over the course of experimental runs. The solid 
line joins the experimental points to aid the eye.
} 
\end{figure*} 
 
An atom periodically kicked in space and time is described by a wave packet  
$\psi (z)$ decomposed into $2\pi$-periodic Bloch states 
$\psi_{\beta}(z)$, that is, 
\begin{equation} 
\psi(z)=\int_0^1d\beta\exp(i\beta z)\psi_{\beta}(z)\, , 
\label{eq:bloch} 
\end{equation} 
where $\beta$ is the quasi-momentum 
(i.e. the fractional part of momentum $p$). 
Quasi-momentum is conserved in the evolution generated by (\ref{eq:ham}), so 
the different Bloch states in (\ref{eq:bloch}) evolve independently of each 
other, whereby their momenta can change only by integers by virtue of the 
kicks. For any given quasi-momentum, the dynamics is formally equivalent to 
that of a rotor (moving on a circle) whose one-period Floquet operator 
is given by 
\begin{equation}  
\Ub \;=\;e^{-{\rm i} k\cos({\hat \theta})}\;e^{-{\rm i}\tau(\Np+\beta)^2/2},  
\label{eq:onecycle}  
\end{equation}  
where $\theta=z$mod$(2\pi)$, and $\Np=-{\rm i}d/d\theta$  
is the angular momentum operator. From 
(\ref{eq:onecycle}) we can immediately 
derive the two necessary conditions for quantum resonant motion: 
if $\tau=2\pi r/q$ ($r,q$ integers)  
then the atomic motion may show asymptotic quadratic growth  
in energy so long as  
$\beta = m/2r$, $0\leq m\leq 2r$, $m$ integer at the same time. Under these conditions 
the Floquet operator (\ref{eq:onecycle}) is also periodic in momentum space, 
with the integer period $q$. As in previous experimental studies  
\cite{QRexp}, we 
focus on the first two fundamental quantum resonances $q=1,2$, for which 
the amplitudes of Bloch waves with $\beta=1/2$ for $q=2$, and $\beta=0,1/2$ 
for $q=1$ at momentum states separated by $q\times 2\hbar k_L$  
exactly rephase after 
each kick. The rephasing condition enforces ballistic propagation of 
the corresponding states in momentum space, so their energy 
grows quadratically in time. The remaining Bloch components of the 
original wave packet (\ref{eq:bloch}),  
with $\beta$ not in the resonant class, 
exchange energy with the kicking laser in a quasiperiodic manner. The 
competition between the resonant and the non-resonant subclasses 
of Bloch states (between ballistic and quasi-periodic propagation) leads 
to {\em linear} growth of the total mean energy, $E\approx k^{2}t/4$, obtained 
by incoherently averaging over the the continuous set of  
quasi-momenta which constitute the atomic ensemble \cite{WGF,WGF2,sandro}. 

\begin{figure*} 
\centering 
\includegraphics[width=10cm]{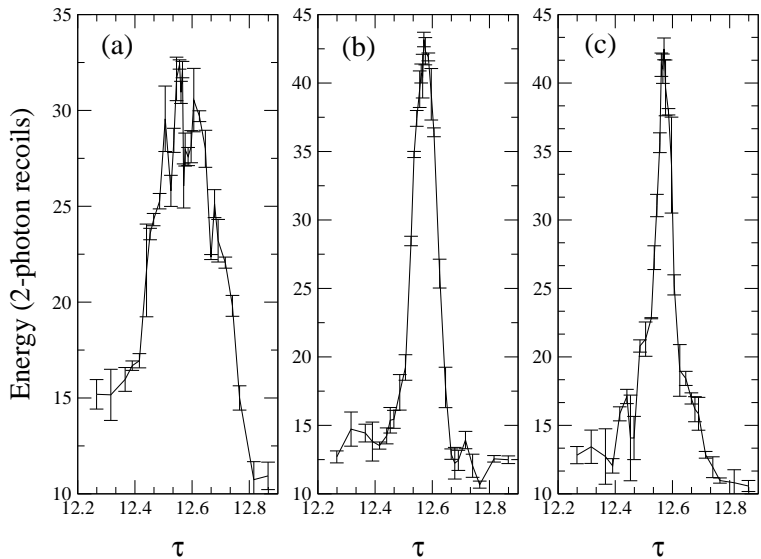} 
\caption{\label{fig:qr4pi} 
Experimentally measured mean energies around  
the second quantum resonance at $\tau=4\pi$ for (a) 5, (b) 10 and (c) 15 
kicks. The kicking strength and  
initial momentum standard deviation were measured to be  
$k=5.0 \pm 0.5$ and $\sigma_p=6.3 \pm 0.1$ respectively. 
Error bars as in Fig. \ref{fig:qr2pi}. We note  
both in this figure and in Fig. \ref{fig:qr2pi} that the resonances  
exhibit some asymmetry, which is thought to be of purely experimental origin
(see the discussion in Section \ref{sthree}).
}
\end{figure*} 
 
For $q=1,2$, we write $\tau=2\pi \ell +\ep$, where $\ep$ denotes the detuning 
from the exact resonance and $l=1,2$. As shown in \cite{WGF,WGF2}, 
the Floquet operator 
(\ref{eq:onecycle}), can then be rewritten as 
\begin{equation} 
\label{eq:onecycleeps} 
 \Ub (t)\;=\;e^{-{\rm i} 
\tilde k\cos({\hat \theta})/|\ep |}\;e^{- {\rm i}\Hb/|\ep |}\;, 
\end{equation} 
with $\tilde k=k|\epsilon|$, $\Ih=|\epsilon|\Np$ as rescaled momentum, and 
\begin{eqnarray} 
\label{eq:epsscal} 
\Hb (\Ih,t) &=& \frac{1}{2} 
\mbox{sign}(\ep) \Ih^2 + \Ih (\pi \ell +\tau \beta)\;. 
\end{eqnarray} 
Introducing the new variables $J=\pm I+\pi\ell+\tau\beta$, 
${\vartheta}=\theta+\pi(1-\mbox{\rm sgn}(\ep))/2$, where $\pm$ denotes the sign  
of $\ep$ $= \mbox{sign}(\ep)$, the quantum evolution can be approximated by 
the $\epsilon$-classical Standard Map derived in \cite{WGF,WGF2,como}
\begin{equation} 
\label{eq:epsmap} 
J_{t+1}=J_t+{\tilde k}\sin (\vartheta_{t+1})\;\;,\;\; 
\vartheta_{t+1}=\vartheta_t+J_t\;, 
\end{equation} 
for ${\tilde k} \ll 1$. $J_t$ implicitly contains the quasi-momentum  
$\beta$, which defines the initial conditions in momentum in the phase space 
generated by the map (\ref{eq:epsmap}) \cite{sandro}. 
 
For small $|\ep|$, the $\ep-$classical dynamics is quasi-integrable, 
and the growth of the energy is dominated by the 
principal $\ep-$classical resonant island around $J=2\pi$ \cite{LL92}.  
The latter island 
is populated only by the values of $\beta$ which are close to the resonant 
ones, whilst the non-resonant quasi-momenta correspond to initial conditions 
outside the nonlinear resonance island \cite{sandro,WGF,WGF2}. 
Moreover, at any time $t$, the ratio between the energy and its value at 
$\ep=0$ is a scaling function of the {\em single} 
variable 
\begin{equation}
x=t\sqrt{k|\ep |}.
\label{eq:xdef}
\end{equation}
The scaling function (which was explicitly 
derived in \cite{WGF,WGF2,sandro}) is 
\begin{equation} 
\label{eq:scal} 
\frac{\langle E_{t,\ep}\rangle} 
{\langle E_{t,0}\rangle} \approx R(x)\equiv 
1-\Phi_0(x)+\frac{4}{\pi x}G(x)\;, 
\end{equation} 
with the functions 
$$ 
\Phi_0(x)\equiv\frac{2}{\pi}\int_0^x 
ds\;\frac{\sin^2(s)}{s^2}\;, 
$$ 
and 
$$ 
G(x) \approx 
\frac{1}{8\pi} 
\int_0^{2\pi}{\rm d}\theta_0 \int_{-2}^{2}{\rm d}J_0 
\overline{J}(x,\theta_0, J_0)^2\;. 
$$ 
$\overline{J}\equiv J/\sqrt{{\tilde k}}$ is the momentum of the pendulum 
approximation to the dynamics generated around the stable fixed point  
of (\ref{eq:epsmap}), rescaled to unit coupling parameter (see 
\cite{WGF,WGF2,sandro} for details).  
 
The one-parameter  
scaling law (\ref{eq:scal}) allows us to deduce the shape and the 
parameter dependence of the resonance peaks elegantly from the experimental 
data, which in the unscaled form is shown in the Figs. \ref{fig:qr2pi} and 
\ref{fig:qr4pi} for $\tau=2\pi$ and $\tau=4\pi$, respectively. 
 
\section{Experimental verification of the scaling law at quantum resonance} 
\label{sthree} 

\begin{figure} 
\centering 
\includegraphics[height=6cm]{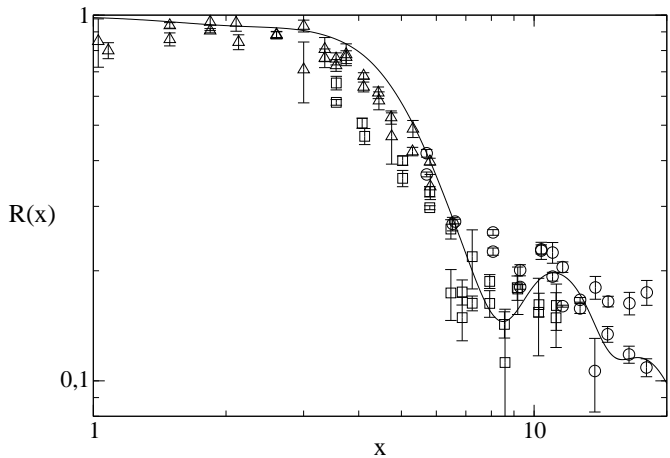} 
\caption{\label{fig:scaling2pi}
Experimental mean energies around $\tau=2\pi$ taken from Fig. \ref{fig:qr2pi} and rescaled as  
$(\langle E_{t,\ep}\rangle-\sigma_p^2/2)/(tk^2/4)$.   
Triangles are for $t=5$, squares for $t=10$ and circles  
for $t=15$. Error bars represent statistical fluctuations over three 
experiments,
and do not take into account fluctuations in $k$ or $\sigma_p$.
The solid line shows the numerically evaluated scaling function
$R(x)$ of Eq. (\ref{eq:scal}).  
We note that, for 10 and 15 kicks, data for  
$|\epsilon | <0.03$ has been omitted due to our inability to accurately 
resolve atomic energies for fast atoms this close to resonance. Experimental 
data for both positive and negative values of $\epsilon$ is plotted.
We would like to note the good correspondence between the
$\epsilon$-classical prediction and the experimental data
for over one order of magnitude in the scaling variable $x$.
} 
\end{figure} 

\begin{figure} 
\centering 
\includegraphics[height=6cm]{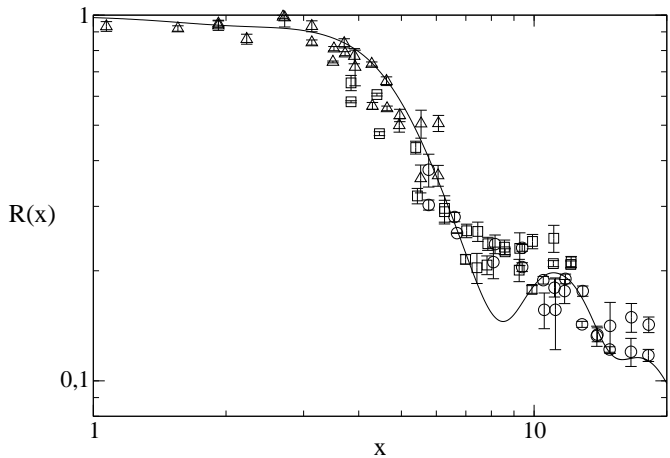} 
\caption{\label{fig:scaling4pi}
Scaled experimental mean energies around $\tau=2\pi$ taken from Fig. \ref{fig:qr4pi};
triangles are for $t=5$, squares for $t=10$ and circles  
for $t=15$ kicks. The solid line 
shows the scaling function $R(x)$ from Eq. (\ref{eq:scal}). 
Again, for 10 and 15 kicks, data too close to resonance, i.e.
for $|\epsilon | < 0.03$, has been omitted. 
} 
\end{figure}

We have used the data obtained for various scans of the mean energy vs.  
the kicking 
period around the quantum resonances $\tau=2\pi$ and $\tau=4\pi$, and for 
kick numbers $t=5,10,15$ to extract the ratio  
$\frac{\langle E_{t,\ep}\rangle} {\langle E_{t,0}\rangle}$. We subtract from 
the numerator the initial energy of the atomic ensemble with 
the characteristic 
width in momentum space $\sigma_p$. The contribution of $\sigma_p^2/2$  
to the energy must 
be subtracted because the derivation of the scaling function $R(x)$ assumed 
an initial atomic momentum distribution in the unit interval $[0,1)$  
\cite{WGF}, corresponding to a uniform distribution of quasi-momenta
$\beta \equiv p_0 \in [0,1)$. Since 
the maximum of the resonance peak  
$\langle E_{t,\ep=0}\rangle$ is experimentally 
the most unstable parameter (due to the early loss of the fastest resonant  
atoms from the experimental detection window 
\cite{QRexp,QRexpnoise,sandro}), we use 
the theoretical value $\langle E_{t,0}\rangle - \sigma_p^2/2
=k^2t/4$ to rescale 
our experimental data, rather than the height of the experimental peak 
itself. Results are presented in Figs.~\ref{fig:scaling2pi} and  
\ref{fig:scaling4pi} for $\tau = 2\pi, 4\pi$ respectively. We see very
 good agreement between the theoretical scaling function $R(x)$
from Eq. (\ref{eq:scal}) 
and our experimental data. Despite the relatively large experimental errors 
due to the uncertainty in the determination of $\sigma_p$
(see discussion below), the data shows the 
characteristic structure, and also the oscillations arising from the 
contribution of the function $G(x)$ at large $x \ge 8$. These oscillations 
arise from the averaged contributions of the initial conditions  
$\overline{J}_0 \in (-2,2)$ within the principal nonlinear resonance island,  
which evolve with different frequencies around the  
corresponding elliptic fixed point of the map (\ref{eq:epsmap}). 
The quasi-momentum classes contributing to $G(x)$ are thus the near-resonant 
values, whilst the non-resonant values contribute to the function  
$1-\Phi_0(x)$, which saturates to a constant for large $x$ 
\cite{WGF,WGF2,sandro}. 
 
We fitted $k$ and $\sigma_p$ for each  
data set and then used these fitted parameters to scale our data. 
In the case of the $\tau=2\pi$ data, the best fit value of $k$ was found to be 
$4.5$ compared to the independently
measured value of $k=4.1 \pm 0.6$. For the $\tau=4\pi$ data, 
the best fit value of $k$ was $5.2$ compared with a measured value of 
$k=5\pm0.5$. The corresponding fitted values of $\sigma_p$ were $5$ and $5.2$ 
two--photon recoils respectively which differ from the  
measured values of $4.53 \pm 0.02$ and $4.3 \pm 0.2$. This difference
is due to the systematic
error involved in determining $\sigma_p$ from the experimental initial
momentum distribution (as discussed in \cite{we}). In particular this
distribution may have noisy exponential wings \cite{Raizen} which
must truncated in order to reliably extract the second moment 
leading to an underestimation of the true initial momentum spread.
 
It is interesting to note that in Figs. \ref{fig:qr2pi} and \ref{fig:qr4pi}, 
there is noticeable asymmetry in the resonance peaks. This degree 
of asymmetry is not predicted by the standard $\epsilon$--classical theory  
and its precise cause has not yet been ascertained. However, the  
asymmetry most likely stems from one or more systematic experimental effects, 
including the effect of small amounts of spontaneous emission ($<0.5\%$ chance 
per kick for the quantum resonance scans) and also from the slightly lesser 
time of flight experienced by atoms for positive as opposed to negative  
$\epsilon$. Asymmetry of the peaks has also been noted in other experiments
probing the structure of the quantum resonances \cite{hoogerland}. 
In any case, this asymmetry does not prevent us from observing 
the structure of the quantum resonances, but leads to a slightly
enhanced scatter of the experimental data points in Figs. \ref{fig:scaling2pi}
and \ref{fig:scaling4pi}.
 
\section{Classical limit of vanishing kicking period} 
\label{sfour} 
 
In spite of the intrinsically quantum nature of the quantum resonances  
as an example of perfectly frequency-matched driving, the method reviewed 
in Section  \ref{theory} allows us to map the quantum dynamics 
onto a purely classical map given by (\ref{eq:epsmap}). 
The latter map is formally equivalent to the usual Standard Map, which 
describes the classical limit of the quantum KR when the kicking period tends 
to zero \cite{Fish}: 
\begin{equation} 
\label{eq:sm} 
J_{t+1}=J_t+{\tilde k}\sin (\theta_{t+1})\;\;,\;\; 
\theta_{t+1}=\theta_t+J_t\;, 
\end{equation} 
now with $J=\tau p= \tau (n+\beta)$, and  
${\tilde k}=k\tau$. Because of the analogy between the maps 
(\ref{eq:epsmap}) and (\ref{eq:sm}), we expect a scaling law for the 
mean energy also in the limit $\tau \to 0$. Since $\tau  \to 0$,  
all quasi-momentum subclasses contribute now similarly to the energy growth, 
and the averaged energy is given only by the initial conditions within  
the principal nonlinear resonance island (see \cite{we} for details): 
\begin{equation} 
\label{eq:bal} 
\langle E_{t,\tau}\rangle \approx  
\tau^{-2}\langle(J_t)^2\rangle/2 \approx k/2\tau G_{cl}(x), 
\end{equation} 
with  
\begin{eqnarray} 
\label{eq:gcl} 
G_{cl}(x) & \equiv & \frac{\sqrt{k}}{2\pi\sqrt{\tau}} 
\int_0^{2\pi}{\rm d}\theta_0 \int_{0}^{\sqrt{\tau/k}}{\rm d}J_0 
\overline{J}(x,\theta_0, J_0)^2 \nonumber \\ 
& \approx& \frac{1}{2\pi}\int_0^{2\pi}{\rm d}\theta_0 \overline{J} 
(x,\theta_0, J_0=0)^2\;, 
\end{eqnarray} 
which depends on the variable $x=t(k\tau)^{1/2}$ (which, given that
$\tau=\epsilon$ for the classical resonance, is the same as the scaling
variable given in Eq. (\ref{eq:xdef}))
and weakly on $k$ and $\tau$, in {\em contrast} 
to the quantum resonant case studied in Section \ref{theory}. 
The dependence of $G_{cl}$ on $\tau$ is negligibly 
small for $\tau \lesssim 1/k$, so that practically, 
$G_{cl}$ can be viewed as a function of the scaling parameter $x$ alone. 
 
For the ratio $\langle E_{t,\tau}\rangle /\langle E_{t,0}\rangle$ 
we then arrive at the scaling function 
\begin{equation} 
\label{eq:scalcl} 
\frac{\langle E_{t,\tau}\rangle }{\langle E_{t,0}\rangle}  
\approx R_{cl}(x) \equiv  \frac{2}{x^2} G_{cl}(x)\;, 
\end{equation} 
which in the limit of vanishing $\tau $ tends to unity, since 
$G_{cl}(x)\approx x^2/2$ for small $x$ \cite{we,sandro}. 
Our result (\ref{eq:bal}) describes \emph{quadratic} growth in mean energy 
as $\tau \to 0$. We note again that in the case  
of quantum resonances, $\epsilon$-classical theory predicts only 
{\em linear} mean energy growth with kick number at quantum  
resonance \cite{WGF,WGF2}. This linear increase is induced 
by the contribution of most quasi-momentum classes which lie  
{\em outside} the classical resonance island. 
For $\tau \to 0$, almost all initial conditions  
(or quasi-momenta) lie {\em within} the principal resonance island, which 
leads to the ballistic growth for the {\em averaged} ensemble energy 
(\ref{eq:bal}).  
 
For finite $\tau>0$ and $t^2k \gg 1/\tau$, we obtain from (\ref{eq:bal}) 
\begin{equation} 
\langle E_{t,\tau > 0}\rangle \approx \frac{k}{2\tau} \alpha\;, 
\label{eq:finite} 
\end{equation} 
since $G_{cl}$ saturates to the value $\alpha \simeq 0.7$ for large $x$. 
Within the stated parameter range, 
this result implies dynamical freezing --  
the ensemble's mean energy is 
independent of kick number. This phenomenon is a classical effect 
in a system with a regular phase space, and has been observed in \cite{we}
for the first time. 
It is distinct from dynamical localization 
which is the quantum suppression of momentum  
diffusion for a chaotic phase space \cite{Casati,Fish}. 
Experimentally, the freezing effect corresponds to the cessation  
of energy absorption from the kicks,  
similar (but different in origin)  
to that which occurs at dynamical localization. 
The freezing may be explained as the averaging over  
all trajectories which start at momenta 
close to zero, and move with different frequencies  
about the principal elliptic fixed point of the map (\ref{eq:sm}). 
 
\begin{figure} 
\centering 
\vspace{0.6cm} 
\includegraphics[height=6cm]{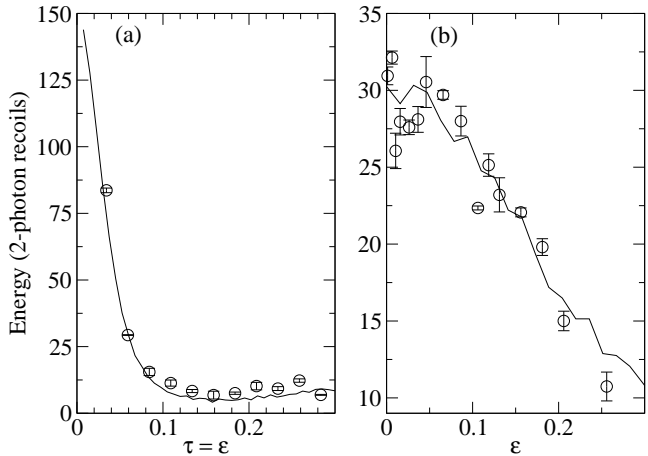} 
\caption{\label{fig:comparet5} 
(a) Circles show experimentally measured mean energies  as  
$\tau \rightarrow 0$ after 5 kicks. 
The measured value of 
$k$ is $4.9 \pm 0.2$. The solid line is classical data for $k=4.9$, as 
generated by the map (\ref{eq:sm}), using practically the
same initial momentum distribution as in the experiment. The 
thermal energy $\sigma_p^2/2$ has 
been subtracted to facilitate comparison with 
the quantum resonance curve in (b). In (b), circles show experimental data 
after 5 kicks near the second quantum resonance  
for positive $\epsilon = \tau-4\pi$ and the experimental parameters are 
as given for Fig. \ref{fig:qr4pi}. The thermal energy $\sigma_p^2/2$ has been  
subtracted. The solid line is $\epsilon$--classical data as generated by 
the map (\ref{eq:epsmap}).
} 
\end{figure}

\begin{figure} 
\centering 
\vspace{0.6cm} 
\includegraphics[height=6cm]{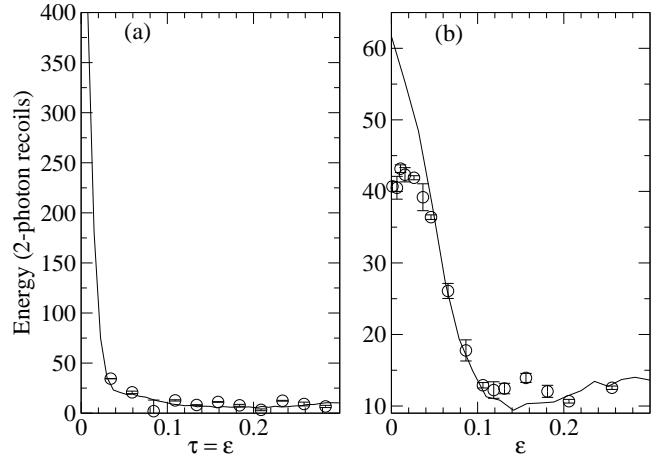} 
\caption{\label{fig:comparet10}(a) Circles show experimental data as 
$\tau \rightarrow 0$ for 10 kicks. The other experimental parameters are the same 
as those given for Fig. \ref{fig:comparet5}(a). The circles in (b) 
 show experimental data once again for the second quantum resonance 
after 10 kicks this time. Other experimental parameters are the same 
as those given for Fig. \ref{fig:comparet5}(b). We note that for the 
quantum resonance in (b), the simulation and experimental results differ most 
markedly near the resonance peak. In this region 
($\epsilon\lesssim 0.03$), some 
fast, resonant atoms are being lost from the experimental viewing 
area leading to a lower energy growth rate than predicted 
theoretically (see discussion in Sections \ref{intro} and \ref{setup}).
Note that in (a), it is not possible to 
probe low values of $\tau = \epsilon$ due to the finite width of the pulses.
} 
\end{figure}

From Eq. (\ref{eq:scalcl}), we immediately see that for the  
``classical'' resonance $\tau \to 0$,  
the resonant peak width scales in time like $(kt^2)^{-1}$, as
at the quantum resonances studied in 
Sections \ref{theory} and \ref{sthree}. However, the tails of
the classical resonance peak decay faster (as $\propto 1/x^2$)
than those at quantum resonance (as $\propto 1/x$, c.f.
 Eq. (\ref{eq:scal})). This very fast 
shrinking of both types of resonance peaks is compared 
in Figs. \ref{fig:comparet5} and \ref{fig:comparet10}.  
 
Both types of these sensitive resonance  
peaks may serve as an experimental tool for determining or calibrating 
parameters in a very precise manner. Additionally, we   
note that the quadratic scaling in time at the quantum resonances and the  
``classical'' resonance, respectively, is much faster and hence much more 
sensitive than the Sub-Fourier resonances detected in a similar context by 
Szriftgizer and co-workers \cite{Lille}. 
A detailed study of the quantum energy  
spectrum of the kicked atoms close to the 
two types of resonances is under way to 
clarify the origin of the observed Sub-Fourier scaling of 
the resonance peaks. 
 
\begin{figure} 
\centering 
\vspace{1.5cm} 
\includegraphics[height=5.5cm]{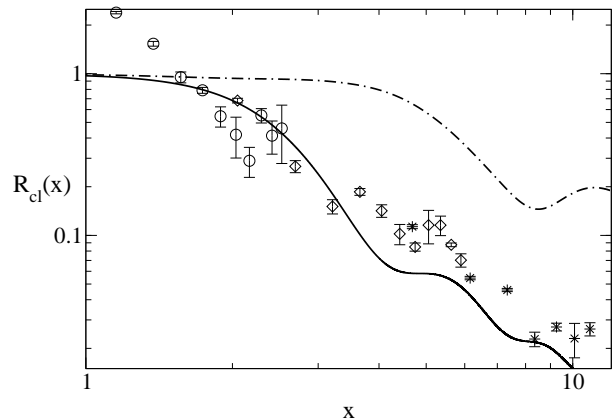} 
\caption{\label{fig:scale_class} 
Rescaled experimental mean energies for 
$\tau = 0.034\ldots 0.284$ (corresponding to $0.33\ldots 2.75\; \rm \mu s$). 
The data is for $k=4.9$ with $t=3$ (circles), 
$t=7$ (diamonds) and $t=16$ (stars). Error bars indicate statistical  
fluctuations over three 
experiments, and do not include variations in $k$ or $\sigma_p$. 
The solid line shows the classical scaling function of Eq.  
(\ref{eq:scalcl}). The dash-dotted line shows the  
scaling function from Eq. (\ref{eq:scal}) (valid for the quantum
resonances) for comparison.
} 
\end{figure}

Finally, we have plotted rescaled experimental data for the $\tau  
\rightarrow 0$ resonance against the scaling function of Eq. 
(\ref{eq:scalcl}),
as seen in Fig. \ref{fig:scale_class}. 
The scaling was performed using the fitted parameters 
as given in Figs. \ref{fig:comparet5} and \ref{fig:comparet10}. 
We note that it is more difficult 
to extract the scaling from experimental data in the classical case, 
as opposed to the quantum case, because the peak of the extremely narrow  
resonance is difficult to probe. This leads to a larger 
uncertainty in the scaled 
energy and the points appear somewhat more scattered than those 
in Figs. \ref{fig:scaling2pi} and \ref {fig:scaling4pi}.
However, the points clearly  
agree much better with the classical scaling function from (\ref{eq:scalcl})
than the $\epsilon$-classical scaling function (\ref{eq:scal})
which is shown in Fig. \ref{fig:scale_class} as a dash-dotted 
line. The clearly different scaling
of the quantum and the ``classical'' resonant
peaks goes along with the same rates at which the
peaks become narrower with time in a Sub-Fourier manner.
 
\section{Conclusions} 
\label{sfive} 

In summary, we have experimentally confirmed a theoretically  
predicted one-parameter scaling law for the resonance peaks in the 
mean energy of a periodically kicked cold atomic ensemble. 
This scaling of the resonant peaks is universal, in the sense 
that it reduces the dependence from all the system's parameters 
to just one combination of such variables. Furthermore, the scaling theory 
works in principle for arbitrary initial momentum distributions. 
In particular, it is valid for the experimentally relevant 
uniformly distributed quasi-momenta at the fundamental 
quantum resonances of the kicked atoms. In the classical limit of 
vanishing kicking period, the dependence on quasi-momentum, as an intrinsic 
quantum variable, disappears entirely, leading to a simpler version of the 
scaling law. The discussed scaling of the experimental data offers one 
the possibility to clearly observe the quantum and ``classical'' 
resonant peak structure over more than one order of magnitude in the scaling
variable. Furthermore, its sensitive 
dependence on the system's parameters may be useful
for high-precision calibration and measurements. 
 
It will be of great interest to clarify whether a similar universal 
scaling law can be found for other time-dependent systems, such as the 
close--to--resonant dynamics of the kicked harmonic oscillator  
\cite{QKHO}, or the driven Harper model \cite{Harpertheo1,Harpertheo2}.  
As with the atom-optics kicked rotor, both of the latter 
systems may be readily realized in laboratory experiments  
\cite{Harperexp,Zoller}.

\section*{Acknowlegements}
M.S. thanks T. Mullins for his assistance in the laboratory prior to these experiments
and acknowledges the support of a Top Achiever Doctoral Scholarship, 03131.

S.W. warmly thanks Prof. Ennio Arimondo and  
PD Dr. Andreas Buchleitner for useful  
discussions and logistical support, and acknowledges funding by the 
Alexander von Humboldt Foundation (Feodor-Lynen Fellowship) and the 
Scuola di Dottorato di G. Galilei della Universit\`a di Pisa.


\begin{thebibliography}{30} 
 
\bibitem{LL92} A.L. Lichtenberg and M.A. Lieberman,  
\textit{Regular and Chaotic Dynamics}, (Springer, Berlin, 1992). 
 
\bibitem{Bayfield} 
J. E. Bayfield,  {\em  
{Quantum Evolution: An Introduction to Time-Dependent Quantum Mechanics}}, 
(Wiley, New-York, 1999). 

\bibitem{Dem} 
W. Demtr\"oder,  {\em  
Laser Spectroscopy: Basic Concepts and Instrumentation}, 
(Springer, Berlin, 2003). 

\bibitem{Casati} 
G. Casati {\em et. al.},  in {\em {Stochastic 
  Behavior in Classical and Quantum Hamiltonian Systems}}, 
  ed. by G. Casati and J. Ford 
  (Springer, Berlin, 1979), p.\ 334. 
 
\bibitem{Izr} F.M. Izrailev and D.L. Shepelyansky  
Sov. Phys. Dokl. \textbf{24}, 996 (1979);  
F.M. Izrailev, Phys. Rep. \textbf{196}, 299 (1990). 
 
\bibitem{QRexp}  
W.H. Oskay {\em et al.}, Opt. Comm. \textbf{179}, 137 (2000); 
M.E.K. Williams {\em et al.}, J. Opt. B: Quantum Semiclass. Opt. 
\textbf{6}, 28 (2004);
G. Duffy \textit{et al.}, Phys. Rev. E \textbf{70}, 056206 (2004).

\bibitem{QRexpnoise} 
M.B. d'Arcy {\em et al.}, Phys. Rev. Lett. \textbf{87}, 074102 (2001);
M.B. d'Arcy {\em et al.}, Phys. Rev. E \textbf{69}, 027201 (2004); 
M. Sadgrove \emph{et al.}, {\em ibid.} \textbf{70}, 036217 (2004).

\bibitem{sandro} 
S. Wimberger,  
Ph.D. Thesis, University of Munich and Universit\`a degli Studi dell' 
Insubria (2004), available at 
http://edoc.ub.uni-muenchen.de/archive/00001687/. 

\bibitem{Pisa}  
S. Wimberger, R. Mannella, O. Morsch, and E. Arimondo,
cond-mat/0501565;  L. Rebuzzini, S. Wimberger, and R. Artuso,
nlin.CD/0410015.

\bibitem{WGF}  
S. Wimberger, I. Guarneri, and S. Fishman,  
Nonlinearity \textbf{16}, 1381 (2003).

\bibitem{WGF2}  
S. Wimberger, I. Guarneri, and S. Fishman,
Phys. Rev. Lett. \textbf{92}, 084102 (2004). 

\bibitem{Fish} 
S. Fishman, in Quantum Chaos, School ``E. Fermi'' CXIX,  
eds. G. Casati \emph{et al.} (IOS, Amsterdam, 1993). 
 
\bibitem{we} 
M. Sadgrove, S. Wimberger, S. Parkins, and R. Leonhardt, 
submitted to Phys. Rev. Lett.

\bibitem{GSZ} 
R. Graham, M. Schlautmann, and P. Zoller, Phys. Rev. A 45, R19 (1992). 

\bibitem{Moore} 
F.L. Moore {\em et al.}, Phys. Rev. Lett. \textbf{75}, 4598 (1995). 

\bibitem{IEEE}  
G. Casati, I. Guarneri, and D. Shepelyansky, 
IEEE J. Quantum Electron. {\bf 24}, 1420 (1988); S. Wimberger and 
A. Buchleitner, J. Phys. A {\bf 34},  7181 (2001). 
 
\bibitem{Chirikov}  
B.V. Chirikov, Phys. Rep. {\bf 52}, 263 (1979). 
 
\bibitem{Monroe1990} C. Monroe, W. Swann, H. Robinson and C. Wieman, 
Phys. Rev. Lett. \textbf{65}, 1571 (1990) 

\bibitem{Raizen}  
B.G. Klappauf, W.H. Oskay, D.A. Steck, and 
M.G. Raizen, Physica D \textbf{131}, 78 (1999).

\bibitem{RTLinuxFAQ} FSMLabs Inc., 
\url{http://www.fsmlabs.com/products/rtlinuxpro/rtlinuxpro_faq.html}. 

\bibitem{Sadgrove2005} M. Sadgrove, T. Mullins, S. Parkins and R. Leonhardt,
Phys. Rev. E \textbf{71}, 027201 (2005).

\bibitem{Fishman}  
R. Bl\"umel, S. Fishman, and U. Smilansky, 
J.~Chem.~Phys. \textbf{84}, 2604 (1986).
 
\bibitem{como}  
S. Fishman, I. Guarneri, and L. Rebuzzini,
J. Stat. Phys. \textbf{110}, 911 (2003);  
Phys. Rev. Lett. \textbf{89}, 084101 (2002).  

\bibitem{hoogerland}
M. Hoogerland, S. Wayper, and W. Simpson, unpublished.

\bibitem{Lille} 
P. Szriftgiser, J. Ringot, D. Delande, and  
J. C. Garreau, Phys.\ Rev.\ Lett.\ {\bf 89}, 224101 (2002); 
H. Lignier, J. C. Garreau, P. Szriftgiser, and D. Delande, 
Europhys. Lett. \textbf{69}, 327 (2005).
 
\bibitem{QKHO} 
G.M. Zaslavsky {\em et al.}, {\em Weak chaos and quasi-regular patterns} 
(Cambridge Univ. Press, 1992); D. Shepelyansky, C. Sire, Europhys. Lett. 
{\bf 20}, 95 (1992); F. Borgonovi and L. Rebuzzini, 
Phys. Rev. E {\bf 52}, 2302 (1995); 
A.R.R. Carvalho and A. Buchleitner, Phys. Rev. Lett. {\bf 93}, 204101 (2004). 
 
\bibitem{Harpertheo1} 
P. Leboeuf, J. Kurchan, M. Feingold, and D. P. Arovas, 
Phys. Rev. Lett. {\bf 65}, 3076 (1990); 
T. Geisel, R. Ketzmerick, and G. Petschel 
{\em ibid.} {\bf 66}, 1651 (1991); 
R. Artuso {\em et. al.}, {\em ibid.} {\bf 69}, 3302 (1992); 
I. Guarneri and F. Borgonovi, 
J. Phys. A {\bf 26}, 119 (1993); 
I. Dana, Phys. Rev. E {\bf 52}, 466 (1995). 
 
\bibitem{Harpertheo2} 
O. Brodier, P. Schlagheck, and D. Ullmo, 
Phys. Rev. Lett. {\bf 87}, 064101 (2001); 
A. R. Kolovsky and H. J. Korsch,  
Phys. Rev. E {\bf 68}, 046202 (2003). 
 
\bibitem{Harperexp} 
H.-J. St\"{o}ckmann, {\em 
Quantum chaos: an introduction\/} (Cambridge University Press, 
Cambridge, 1999); T.M. Fromhold {\em et. al.}, Nature {\bf 428}, 726 (2004). 
 
\bibitem{Zoller} 
S.A. Gardiner, J.I. Cirac, and P. Zoller, 
Phys. Rev. Lett. {\bf 79}, 4790 (1997); 
S.A. Gardiner {\em et. al.}, Phys.~Rev.~A {\bf 62}, 023612 (2000). 
 
 
\end{thebibliography}
\end{document}